

\def\pmb#1{\setbox0=\hbox{$#1$}%
\kern-.025em\copy0\kern-\wd0
\kern.05em\copy0\kern-\wd0
\kern-.025em\raise.0433em\box0 }
\font\cs=cmcsc10
\vsize=7.5in
\hsize=5.6in
\tolerance 10000

\baselineskip 12pt plus 1pt minus 1pt
\pageno=0
\centerline{\bf FINITE AND INFINITE SYMMETRIES IN}
\smallskip
\centerline{{\bf (2+1)-DIMENSIONAL FIELD THEORY}\footnote{*}{This
work is supported in part by funds
provided by the U. S. Department of Energy (D.O.E.) under contract
\#DE-AC02-76ER03069 (RJ), and \#DE-AC02-89ER40509(S-YP).}}
\vskip 24pt
\centerline{R. Jackiw}
\vskip 12pt
\centerline{\it Center for Theoretical Physics}
\centerline{\it Laboratory for Nuclear Science}
\centerline{\it and Department of Physics}
\centerline{\it Massachusetts Institute of Technology}
\centerline{\it Cambridge, Massachusetts\ \ 02139\ \ \ U.S.A.}
\vskip 12pt
\centerline{and}
\vskip 12pt
\centerline{So-Young Pi}
\vskip 12pt
\centerline{\it Physics Department}
\centerline{\it Boston University}
\centerline{\it 350 Commonwealth Avenue}
\centerline{\it Boston, Massachusetts\ \ 02115}
\vfill
\centerline{Dedicated to Franco Iachello on his 50th Birthday}
\vfill
\noindent Recent Problems in Mathematical Physics, Salamanca, Spain, June
1992; XIX International Colloquium on Group Theoretical Methods in Physics,
Salamanca, Spain, July 1992, Condensed Matter and High Energy Physics,
Cagliari, Italy, September 1992.
\vskip .3in
\centerline{ Typeset in $\TeX$ by Roger L. Gilson}
\vskip -12pt
\noindent CTP\#2110\hfill June 1992

\noindent BU HEP~92-21
\eject
\baselineskip 24pt plus 2pt minus 2pt
\centerline{\bf ABSTRACT}
\medskip
These days, Franco Iachello is {\it the\/} eminent practitioner applying
classical and finite groups to physics.  In this he is following a
tradition at Yale, established by the late Feza Gursey, and succeeding Gursey
in the Gibbs chair;  Gursey in turn, had Pauli as a mentor.  Iachello's
striking achievement has been to find an actual realization of arcane
supersymmetry within mundane
adjacent even-odd nuclei.  Thus far this is the only {\it
physical\/} use of supersymmetry, and its fans surely must be surprised
at the venue.   Here we describe the role of $SO(2,1)$ conformal
symmetry in non-relativistic Chern--Simons theory:  how it acts, how it
controls the nature of solutions, how it expands to an infinite group on the
manifold of static solutions thereby rendering the static problem completely
integrable.  Since Iachello has also used the $SO(2,1)$ group
in various contexts, this
essay is presented to him on the occasion of his fiftieth birthday.
\vfill
\eject
\noindent{\bf I.\quad INTRODUCTION}
\medskip
\nobreak
We shall discuss finite- and infinite-dimensional conformal symmetries of {\it
field theories with non-relativistic kinematics\/}.  Such field theories
also describe the second quantization of {\it non-relativistic particle
mechanics\/}.  Particle mechanics, with its second order  in time
dynamics, has the structure of a {\it relativistic field theory\/} in one time
and zero space dimensions, and a relativistic field theory in any dimension
can enjoy conformal symmetry.  Thus there are family relationships between the
conformal symmetries of non-relativistic field theory, non-relativistic
particle mechanics and relativistic field theory, and our first task is to
describe these interrelations.

A conformal transformation in $(D+1)$-dimensional relativistic field theory
changes the {\it independent\/}
variables, {\it viz.\/} the space-time coordinates
$x^\mu$ of the fields (fields are {\it dependent\/} variables),
and infinitesimally reads
$$\delta_f x^\mu = - f^\mu(x) \eqno(1.1)$$
where $f^\mu$ is a {\it conformal Killing vector\/}, {\it i.e.\/} $f^\mu$
satisfies the {\it conformal Killing equation\/}.
$$\partial_\mu f_\nu + \partial_\nu f_\mu = {2\over D+1} g_{\mu\nu}
\partial_\alpha f^\alpha \eqno(1.2)$$
Here $g_{\mu\nu}$ is the Minkowski metric tensor with signature
$(1,-1,-1,\ldots)$ and $D$ is the spatial dimensionality.

As is well-known,
Eq.~(1.2) has the {\it finite\/} number of
${1\over 2}(D+2)(D+3)$ solutions for $D>1$, and conformal
transformations form an $SO(2,D+1)$ group.  The solutions to (1.2) comprise
$$\eqalignno{
\hbox{$D+1$ space-time translations}\hskip .3in
f^\mu(x) &= a^\mu \ \ ,\qquad a^\mu\ \hbox{constant} &
(\hbox{1.3a})\cr
\hbox{${1\over 2} D(D+1)$ space-time rotations}\hskip .3in
f^\mu(x) &= \omega^\mu{}_\nu x^\nu\ \  ,\qquad \omega_{\mu\nu} =
-\omega_{\nu\mu} &(\hbox{1.3b})\cr
\hbox{a single scale transformation}\hskip .3in
f^\mu(x) &= ax^\mu\ \ ,\qquad a\ \hbox{constant}&(\hbox{1.3c})\cr
\hbox{$D+1$ special conformal transformations}\hskip .3in
f^\mu(x) &= 2c\cdot x\, x^\mu - c^\mu x^2 \ \ ,\qquad c^\mu\
\hbox{constant} \cr&&(\hbox{1.3d})\cr}$$
The finite versions of these are, respectively,
$$\eqalignno{x^\mu &\to x^\mu + a^\mu &(\hbox{1.4a})\cr
x^\mu &\to \Lambda^\mu{}_\nu x^\mu\ \ ,\qquad \Lambda^\mu{}_\alpha
\Lambda^\nu{}_\beta g_{\mu\nu} = g_{\alpha\beta}&(\hbox{1.4b}) \cr
x^\mu&\to e^a x^\mu &(\hbox{1.4c}) \cr
x^\mu&\to {x^\mu - c^\mu x^2\over 1-2c\cdot x + c^2 x^2} &(\hbox{1.4d}) \cr}$$
The last, the finite special conformal transformation, can also be seen as an
inversion, $x^\mu\to x^\mu/x^2$, followed by a translation and another
inversion, {\it i.e.\/} a translation in the inverted coordinate.

At $D=1$ there exists an {\it infinite\/} number of solutions to (1.2)
corresponding to arbitrary redefinition of $x^\pm = {1\over\sqrt{2}}
(x^0\pm x^1)$
and forming an infinite parameter group.  Infinitesimally we have
$$\delta_f x^\pm = - f^\pm (x^\pm)\ \ ,\qquad f^\pm\ \hbox{arbitrary}
\eqno(1.5)$$
while the finite version reads
$$x^\pm\to X^\pm (x^\pm) \ \ ,\qquad X^\pm\ \hbox{arbitrary}\eqno(1.6)$$

A linear conformal transformation on a space-time multiplet of Lorentz
covariant relativistic fields $\varphi$,
{\it i.e.\/} on the dependent variables,
can be taken as
$$\delta_f\varphi = f^\alpha \partial_\alpha\varphi
+ \partial_\alpha f_\beta \left(
{\Delta\over D+1} g^{\alpha\beta} + {1\over 2}
\Sigma^{\alpha\beta}\right)\varphi \eqno(1.7)$$
Here $\Sigma^{\alpha\beta}$ is the spin-matrix, acting on the space-time
components of $\varphi$
and $\Delta$ is a constant, called the scale-dimension of
$\varphi$.  When the Lagrange density for $\varphi$ possesses a conventional
relativistic kinetic term --- quadratic in derivatives for Bose fields, linear
for Fermi fields --- the kinetic action is invariant against conformal
transformations (1.7) provided
$$\eqalignno{\Delta &= {D-1\over 2} & \hbox{bosons} \hskip 1in\hbox{(1.8a)} \cr
\Delta &= {D\over 2} & \hbox{fermions} \hskip 1in\hbox{(1.8b)}\cr}$$
[These values for $\Delta$ correspond to the dimensionality of a field in
units of inverse length when
$\hbar$ and $c$ are scaled to unity.]  Also the Bose field monomial
$${\cal L}_I = \varphi^{2\left( {D+1\over D-1}\right)} \eqno(1.9)$$
leads to an invariant action $\int d^{D+1}x{\cal L}_I$.

At $D=1$, Bose fields become dimensionless, see (1.8a),  and the
conformally monomial (1.9) cannot be formed.
Nevertheless, there exists a non-trivial
conformally invariant theory --- the completely integrable Liouville theory,
$${\cal L}_{\rm Liouville} = {1\over 2} \partial_\mu \varphi \partial^\mu
\varphi - {\mu^2\over\beta^2} e^{\beta\varphi}\eqno(1.10)$$
whose action is invariant provided the single-component
scalar field $\varphi$ is transformed according
to an inhomogenous generalization of (1.7),
$$\delta_f \varphi = f^\alpha \partial_\alpha\varphi + {1\over \beta}
\partial_\alpha f^\alpha\eqno(\hbox{1.11a})$$
or equivalently
$$\delta_f\, e^{\beta\varphi} = \partial_\alpha \left( f^\alpha
e^{\beta\varphi}\right) \eqno(\hbox{1.11b})$$

The hallmark of a conformally invariant theory is that an energy-momentum
tensor $T^{\mu\nu}$ can be constructed, which is conserved (as a
consequence of translation invariance) symmetric (as a consequence of Lorentz
invariance) and traceless (as a consequence of conformal invariance).  Thus
conformal invariance in a relativistic field may be summarized by a relation
between the energy density ${\cal E} = T^{00}$ and the trace of the spatial
stress tensor $T^{ij}$.
$${\cal E} = \sum\limits^D_{i=1} T^{ii}\eqno(1.12)$$
The currents $j^\mu_f$ that are conserved as a consequence of conformal
invariance are then constructed in the Bessel--Hagen
form from a projection of $T^{\mu\nu}$ on the conformal Killing vector,
$$j^\mu_f = T^{\mu\nu} f_\nu\eqno(\hbox{1.13a})$$
and the constants of motion read
$$C_f = \int d^D  r \left( {\cal E} f^0 - \pmb{\cal P}\cdot {\bf f}
\right) \eqno(\hbox{1.13b})$$
where $\pmb{\cal P}$ is the momentum density.
$${\cal P}^i = T^{0i}\eqno(1.14)$$
[Frequently it is necessary to ``improve'' the energy-momentum tensor obtained
by Noether's theorem or by general relativistic considerations.]

The kinetic Lagrangian for non-relativistic motion of point-particles in
$d$-dimensional space is quadratic in derivatives with respect to time, which
is the single independent variable.  Hence it has the structure of a
$(0+1)$-dimensional relativistic ``field theory,'' where the ``field'' is
particle position ${\bf r}(t)$, now a dependent variable,
while the $d$-spatial dimensions form an ``internal'' space and ${\bf r}$ is a
vector in this space.
``Conformal'' transformations degenerate into reparametrization of the single
independent variable, {\it i.e.\/} time.  The
previous discussion can be taken at $D=0$, but the conformal Killing equation
(1.2) becomes vacuous.  Nevertheless, one easily shows that (1.3) and (1.4)
[with (1.3b) and (1.4b) absent] are invariances of the kinetic term
provided the dependent variable ${\bf r}$ transforms
according to
$$\delta{\bf r}  = f \dot{\bf r} - {1\over 2} \dot f  {\bf r} \eqno(1.15)$$
when the independent variable $t$ changes by
$$\eqalignno{\delta t &= -f(t) &(\hbox{1.16a}) \cr
f(t) &= a,at,at^2 &(\hbox{1.16b})\cr}$$

These comprise the $D=0$ restriction of (1.7) and (1.8a) with $\Delta = -
1/2$, and form an $SO(2,1)$ group of transformations.  It is seen that
${\bf r}$ has scale dimension of $-1/2$, {\it i.e.\/}
it scales as $\sqrt{t}$ --- this is a consequence of having scaled $\hbar$ to
unity and with non-relativistic kinematics one can take $m$ to be
dimensionless.
Further from (1.9) at $D=0$ one sees that the $r^{-2}$ potential also gives an
invariant action since $\dot r^2$ and $1/r^2$, the two terms comprising a
Lagrangian or Hamiltonian, scale in the same way. When
$$H = {1\over 2} m\dot r^2 + {\lambda\over r^2}\ \ .\eqno(1.17)$$
the three constants of motion
$$C_f = Hf - {m\over 4} \left( {\bf r} \cdot \dot{\bf r} + \dot{\bf r}
\cdot {\bf
r}\right) \dot f + {m\over 4} r^2 \ddot f \eqno(1.18)$$
also generate the transformation (1.14)
when the canonical momentum ${\bf p}$, conjugate to ${\bf r}$, is taken to be
$m\dot{\bf r}$, and their algebra realizes the $SO(2,1)$ Lie algebra.  One can
view (1.18) as the one-time, zero-space
analog of the Bessel--Hagen expression (1.13).

In specific spatial dimensions $d$, other interactions in addition to the
$1/r^2$ potential preserve
conformal invariance.  Examples for $d=3$ and $2$ are, respectively,
interaction with a Dirac magnetic monopole and a point vortex.  For these
(1.17) and (1.18) retain the same form, but the relation between canonical
momentum and velocity is modified by the presence of a vector potential
$${\bf p} = m\dot{\bf r} + {\bf A} \eqno(1.19)$$
where at $d=3$, ${\bf A}$ is the Dirac vector potential that gives rise to a
monopole magnetic field of strength $g_m$
$${\bf A} = {\bf A}_D\ \ ,\qquad
{\bf B} = \pmb{\nabla}\times {\bf A}_D = {g_m{\bf r}\over r^3}\eqno(1.20)$$
while in planar physics at $d=2$, ${\bf A}$ is the vortex potential,
$${\bf A} = {\Phi\over 2\pi} \pmb{\nabla}\theta\ \ ,\qquad {\bf B} =
\pmb{\nabla}\times {\bf A} = \Phi \delta^2 ({\bf r}) \eqno(1.21)$$
with $\tan\theta = y/x$ and $\Phi$ being the flux of the vortex.  [In the
above we use the amusing distributional formula $\pmb{\nabla}\times
\pmb{\nabla}\theta = 2\pi\,\delta^2({\bf r})$.]

Furthermore, at $d=2$, the $\delta$-function potential also scales as
$r^{-2}$, hence it also appears to be conformally invariant.  In the
following, we shall consider the two-dimensional particle model, with a vortex
(1.21) and $\delta$-function interactions; {\it i.e.\/} the Hamiltonian is
$$H = {1\over 2} m\dot r^2 - g\, \delta^2({\bf r}) = {1\over 2m} ({\bf p} -
{\bf
A})^2 - g \,\delta^2({\bf r}) \eqno(1.22)$$

Non-relativistic particle quantum mechanics may be second quantized, and in
this way one is led to a non-relativistic quantum field theory, with
field-theoretic symmetries that encode the above $SO(2,1)$ particle symmetries,
but now realized in an action on the dependent field variable $\psi$, which is
a function of the independent variables $t$, and ${\bf r}$, where ${\bf r}$ is
a two-dimensional vector.

In these lectures, we shall explain these symmetries of non-relativistic field
theory, at $d=2$ --- planar physics. Also we shall
show how the $SO(2,1)$ group can expand to an infinite-dimensional group of
conformal reparametrizations of the two-dimensional spatial plane.

Contexts, wherein recently
there is encountered a non-relativistic, planar field
theory, are the following two:
\medskip
\item{1)}{\cs Second quantized,
non-relativistic particles with Abelian or non-Abelian charge,
interacting with a gauge field whose kinetic dynamics is provided by the
Chern--Simons action.}  The field theoretic action in the Abelian case, with
gauge potentials eliminated in terms of matter variables, is
$$I = \int dt\, d^2r \left\{ i \psi^*\partial_t \psi - {1\over 2m} \left|{\bf
D}\psi\right|^2 + {g\over 2} \rho^2\right\} \eqno(1.23)$$
where the covariant derivative ${\bf D}$ involves a gauge potential ${\bf A}$
$${\bf D} = \pmb{\nabla} - i {\bf A}\eqno(1.24)$$
which is determined by the matter density $\rho= |\psi|^2$
$${\bf A}(t,{\bf r}) = \pmb{\nabla}\times {1\over 2\pi\kappa} \int d^2r'
\ln\left| {\bf r} - {\bf r}'\right| \rho(t, {\bf r}') \eqno(1.25)$$
so that the Chern--Simons Gauss law is satisfied.
$$B = \pmb{\nabla}\times {\bf A} = - {1\over\kappa}\rho\eqno(1.26)$$
Here $1/\kappa$ measures the interaction strength and without
loss of generality we may take it to be non-negative.  [In the plane, the
cross product of two vectors defines a scalar, and cross multiplication with a
single vector results again in a vector; in components: $s=\epsilon^{ij}
v^i_{(1)}v^j_{(2)}$; $v^i_{(1)} = \epsilon^{ij} v^j_{(2)}$.]  Also there is
present in (1.23) a quartic self-interaction of strength $g$, which is the
second-quantized description of a two-body $\delta$-function interaction.
Thus (1.23) provides the second quantization of (1.22); it may also be
presented as
$$I = \int dt\, d^2r\left\{ i \psi^* \partial_t \psi - {\cal E}\right\}
\eqno(1.27)$$
where the energy density is given by the formula$^1$
$${\cal E} = {1\over 2m} \left| {\bf D}\psi\right|^2 - {g\over 2}\rho^2
\eqno(1.28)$$
\item{2.}{\cs
Effective action for gravity or Abelian vector gauge theories in the eikonal
(large-$s$, fixed-$t$) limit.}  It is found that in the eikonal regime, the
conventional action [Einstein--Hilbert for gravity, Maxwell for gauge
theory] can be written as a total derivative on a two-dimensional space-time
plane imbedded in four-dimensional space-time.  By integrating the total
derivative onto a curve (parametrized by $\tau$) forming the boundary of
that two-dimensional plane, the action [without sources] becomes
$$I_{\rm eikonal} = {1\over 2} \int d\tau\, d^2r\left( \partial_i \Omega^+
\partial_i\dot\Omega^--\partial_i \Omega^- \partial_i\dot\Omega^+ \right)
\eqno(1.29)$$
where the overdot denotes differentiation with respect to $\tau$, while
${\bf r}$ and $\partial_i$, $i=1,2$, refer to the remaining two spatial
directions, and $\Omega^\pm$ are the surviving field (gravitational, vector)
degrees of freedom.$^2$  Upon defining
$$\psi = {1\over \sqrt{2}}\left( \partial_x+i\partial_y\right) \left( \Omega^+
- i \Omega^-\right) \eqno(1.30)$$
(1.29) may be rewritten, apart from total derivative contributions, as
$$I_{\rm eikonal} = \int d\tau\,d^2r\, i \psi^* \partial_\tau \psi
\eqno(1.31)$$
Precisely the same form as (1.27) is revealed, except now the
energy density vanishes.
\goodbreak
\bigskip
\noindent{\bf II.\quad SYMMETRIES}
\medskip
\nobreak
The field theoretic Lagrangian
$$\eqalignno{L &= \int d^2ri\, \psi^*\partial_t \psi - H &(\hbox{2.1a}) \cr
H &= \int d^2r\,{\cal E} &(\hbox{2.1b})\cr}$$
represents both the non-relativistic Chern--Simons model (1.27), (1.28) and the
eikonal limits of relativistic theory (1.29), (1.31), where in the latter
case ${\cal E}$ vanishes.  So we use (2.1) as the basis for our discussion of
symmetries in both cases, keeping in mind that the vanishing of ${\cal E}$
for the latter, renders much of
the analysis vacuous, but as we shall see, not without
relevance to the former.  Throughout we shall solely deal with the Abelian
Chern--Simons theory, though as far as symmetry properties are concerned, the
non-Abelian model behaves similarly.  Also discussion is confined to classical
symmetries of the field theory, viewed as a classical non-linear system.
Anomalies in the symmetries due to quantum effects will only be mentioned in
the Conclusion.

The energy density (1.28) of the Chern--Simons model
$${\cal E} = {1\over 2m} \left|{\bf D}\psi\right|^2 - {g\over 2}\rho^2
\eqno(\hbox{2.2a})$$
is identically equal to
$${\cal E} = {1\over 2m} \left| D\psi\right|^2 - {1\over 2} \left(
\pmb{\nabla}\times {\bf j} + {1\over m} B\rho\right) - {g\over 2} \rho^2
\eqno(\hbox{2.2b})$$
where the current ${\bf j}$ is
$${\bf j} = {1\over m} \Im \psi^*{\bf D}\psi\eqno(2.3)$$
and $D$ is the holomorphic, gauge covariant derivative.
$$D\equiv D_x -iD_y\eqno(2.4)$$
The curl of ${\bf j}$ will
not contribute to a variational derivation of the equations of motion, nor
will it contribute to the integrated total energy, provided the current is
sufficiently well-behaved at the edge of space [which lies at infinity].  With
the assumption of requisite regularity for ${\bf j}$ and the use of the
Chern--Simons Gauss law constraint (1.26), the energy/Hamiltonian may be
presented as
$$E = H = \int d^2{\bf r} {\cal H}\ \ ,\qquad {\cal H} = {1\over 2m} \left|
D\psi\right|^2 - {1\over 2} \left( g - {1\over m\kappa}\right) \rho^2
\eqno(2.5)$$
This is the form of the Hamiltonian density that we shall scrutinize as regards
to
the symmetries of the model.

The symmetries are of two kinds: a) symmetries of the
action, {\it i.e.\/} transformations which leave the action invariant and lead
to constants of motion by Noether's theorem --- these are well-known and
include the obvious Galileo transformations, and the $SO(2,1)$
time-reparametrization conformal symmetries specific to the planar model with
which we are here concerned; and b) symmetries of the critical points of the
action, {\it i.e.\/} transformations which leave selected equations of motion
invariant, map solutions into solutions, but do not give rise to constants of
motion because they do not leave invariant the action away from its critical
points.  These symmetries of the Chern--Simons model
have not been previously studied systematically,
though their occurrence in static solutions at $g = 1/m\kappa$
had been noted: they comprise conformal reparametrization symmetries of the
two-dimensional plane.$^{3,\,4}$
\goodbreak
\bigskip
\noindent{\bf II.A.\quad Finite-Dimensional Symmetry Group of the Action}
\medskip
\nobreak
As befits any respectable field theory, our model is invariant against time
translation, space translation and rotation, as well as against Galileo
boosts because dynamics is non-relativistic.  The last invariance is perhaps
unexpected in the presence of gauge fields, which conventionally are invariant
against the {\it Lorentz\/} boosts of {\it special relativity\/}
(indeed this led to the
invention of special relativity!).  What distinguishes the present situation is
that the gauge dynamics are of the Chern--Simons variety, and the
Chern--Simons term, being topological, is invariant against {\it all\/}
space-time transformation, while the non-relativistic
matter system is only Galileo invariant.
 The transformation laws on the fields are familiar.  For the first three,
$$\eqalignno{\hbox{time translations}\hskip .5in t' &= t + a &(2.6)\cr
{\bf r}' &= {\bf r}\cr\noalign{\vskip 0.2cm}
\hbox{space translation}\hskip .5in t'&= t &(2.7) \cr
{\bf r}' &= {\bf r} + {\bf a} \cr\noalign{\vskip 0.2cm}
\hbox{space rotation}\hskip .5in t' &=t&(2.8) \cr
r'{}^{i}&= R^{ij} (\omega)r^j \cr}$$
[$R^{ij}(\omega)$ is the rotation matrix through angle $\omega$]  the field
transforms as a scalar.
$$\psi'(t',{\bf r}')=\psi(t,{\bf r})\eqno(2.9)$$
The Galileo transform
$$\eqalign{\hbox{Galileo boost}\hskip .5in t' &= t\ \ ,\cr
{\bf r}' &= {\bf r} + {\bf v}t \cr}\eqno(2.10)$$
requires a 1-cocycle in the field transformation law.
$$\psi'(t',{\bf r}') = e^{im{\bf v}\cdot\left( {\bf r} + {1\over 2} {\bf v}
t\right)} \psi(t,{\bf r}) \eqno(2.11)$$

Additionally, our system is invariant against conformal reparametrizations of
time.  These include three $SO(2,1)$ transformations of time: translation (2.6)
and (2.9),
$$\eqalign{\hbox{time dilation}\hskip .5in
t'&=at\cr{\bf r}'&= \sqrt{a}\,{\bf r} \cr}\eqno(2.12)$$
for which the field transformation law acquires a weight factor,
$$\psi' (t',{\bf r}') = {1\over \sqrt{a}} \psi(t,{\bf r}) \eqno(2.13)$$
and translation of inverse time,
$$\eqalign{\matrix{\hbox{conformal time}\cr \hbox{transformation}\cr}\hskip
.5in
{1\over t'} &= {1\over t}+a \cr
{\bf r}' &= {1\over 1+at}{\bf r} \cr}\eqno(2.14)$$
where the field transformation law has both a weight factor and 1-cocycle.
$$\psi'\left( t',{\bf r}'\right) =\left( 1+at\right) e^{-imar^2\over 2(1+at)}
\psi(t,{\bf r}) \eqno(2.15)$$

One can check that owing to the weight factors, which are square roots of the
Jacobian, the density $\rho$ transforms with the Jacobian, $J$
$$\eqalignno{\rho'\left( t',{\bf r}'\right) &= J \rho(t,{\bf r}) &(2.16)\cr
J&\equiv \det\left\{ {\partial r^i\over \partial r'{}^j}\right\} &(2.17)\cr}$$
and the vector potential, defined by (1.25), transforms covariantly.
$$A'{}^i \left( t',{\bf r}'\right) = A^j (t,{\bf r}) {\partial r^j\over
\partial r'{}^i} \eqno(2.18)$$
The action is invariant, and the conserved generators can be obtained from
Noether's theorem.

Alternatively one records the formula for the energy momentum tensor
components:
$$\eqalignno{\hbox{energy density}\hskip .5in &T^{00} \equiv {\cal E}
= {1\over 2m} \left| {\bf D}\psi \right|^2 -
{g\over 2} \rho^2 &(2.19)\cr
\hbox{momentum density}\hskip .5in
&\pmb{\cal P} = m{\bf j} = \Im\psi^*{\bf D}\psi
&(2.20)\cr}$$
These satisfy continuity equations with energy flux ${\bf T}$,
$$\hbox{energy flux}\hskip .5in
{\bf T} = - {1\over 2} \left( \left( D_t \psi\right)^* {\bf D} \psi +
\left( {\bf D}\psi\right)^* D_t \psi\right)\eqno(2.21)$$
and momentum flux --- the stress tensor $T^{ij}$.
$$\eqalign{\hbox{momentum flux}\hskip .5in
T^{ij} &= {1\over 2} \left( \left( D_i\psi\right)^* \left(
D_j\psi\right) + \left( D_j \psi\right)^* \left( D_i\psi\right) - \delta^{ij}
\left( D_\kappa\psi\right)^* \left( D_\kappa\psi\right)\right) \cr
&\quad + {1\over 4} \left( \delta^{ij}\nabla^2 - 2\partial_i \partial_j\right)
\rho + \delta^{ij} {\cal E} \cr}\eqno(2.22)$$
Here $D_t = \partial_t + iA^0$, where $A^0$ solves the Chern--Simons equation
that supplements (1.26).
$$A^0 (t,{\bf r}) = - {1\over 2\pi\kappa} \int d^2r'\, \epsilon^{ij}
{\left( r^i - r'{}^i\right) \over \left| {\bf r} - {\bf r}'\right|^2} j^j
(t,{\bf r}') \eqno(2.23)$$
The continuity equations read
$$\eqalignno{\partial_t {\cal E} + \pmb{\nabla}\cdot {\bf T} &= 0 &(2.24) \cr
\partial_t {\cal P}^i + \partial_j T^{ij} &= 0 &(2.25) \cr}$$

Note that energy flux ${\bf T}$
does not equal momentum density $\pmb{\cal P}$, since our theory is not
Lorentz invariant.  But it is rotationally invariant; that is why the
stress-tensor is symmetric in its spatial indices.  Also $T^{ij}$ satisfies
$$2{\cal E} = \sum^2_{i=1} T^{ii} \eqno(2.26)$$
and this reflects the $SO(2,1)$ invariance, being the non-relativistic analog
of (1.12).

Of course, the theory is also phase invariant; this produces one more
continuity
equation
$$\partial_t\rho + \pmb{\nabla}\cdot {\bf j} = 0 \eqno(2.27)$$
where the proportionality of the matter flux current ${\bf j}$ to the
momentum density (2.20) is a consequence of Galileo invariance.

The constants of motion are now constructed from moments of the energy
momentum tensor and $\rho$.  They are, respectively
$$\eqalignno{ \hbox{energy} \hskip .5in
E &= \int d^2 r \,{\cal E} &(2.28) \cr
\hbox{momentum}\hskip .5in  {\bf P} &= \int d^2r\,\pmb{\cal P} &(2.29) \cr
\hbox{angular momentum}\hskip .5in M &= \int d^2r\,{\bf r}\times \pmb{\cal P}
&(2.30)\cr
\hbox{Galileo boost}\hskip .5in
{\bf B} &= t{\bf P}-m \int d^2r\,{\bf r}\rho &(2.31)\cr
\hbox{dilation}\hskip .5in D &= tE - {1\over 2}
\int d^2r\,{\bf r}\cdot\pmb{\cal P}
&(2.32) \cr
\hbox{special conformal} \hskip .5in
K &= - t^2 E + 2tD + {m\over 2} \int dr^2\,r^2\rho &(2.33) \cr
\hbox{matter number}\hskip .5in
N &= \int d^2r\,\rho &(2.34) \cr}$$
It is straightforward to verify that all these are conserved, as a consequence
of the continuity equations and the special properties (symmetry and trace)
of the stress tensor.  Note that collectively the constants may be
written analogously to (1.18) and to the Bessel--Hagen expression (1.13), as
$$C_f = \int d^2r\,{\cal E}f_1 - \int d^2r\,\pmb{\cal P}f_2 +
\int d^2r\,\rho f_3\eqno(2.35)$$
for suitable $f_i$.

The above transformations may be generalized in the following interesting
manner.  One may consider an {\it arbitrary\/} reparametrization of time
[rather than the specific forms (2.6), (2.12), (2.14)].  Also one may shift
${\bf r}$ by a vector with arbitrary time dependence [rather than constant
(2.7) or linear (2.10) in time].  Finally, one may rotate ${\bf r}$ as in
(2.8), but with a time-dependent angle.  Of course, these transformations are
no longer symmetry operations of our theory, but they map our model onto
another, closely related one: it is found that the above transformations
introduce interactions with external fields.  Specifically, after these
transformations are carried out there arise external electric and magnetic
fields, determined by the parameters of the transformations, hence
the fields are time-dependent
but constant in space; additionally there is an external
harmonic potential, with time-varying frequency.$^5$  [With specific
time-dependence, the parameters can conspire to produce static electric and
magnetic fields as well as time-independent harmonic forces; also one can
suppress selectively any of the external effects.]

Higher symmetries are widely studied these days in field theory, but it seems
that rarely do they provide specific dynamical information about a model ---
rather they give an elegant frame for describing solutions and other
properties.

As an exception that proves the rule, we now show that the conformal
symmetries allow deriving the following useful result about the highly
non-linear dynamics of our Chern--Simons theory: all static
solutions carry zero energy.$^6$  This follows immediately from (2.32) and/or
(2.33):  the left sides are time independent, and so are the last terms in the
right sides for time independent $\rho$ and ${\cal P}=m{\bf j}$, {\it i.e.\/}
for static solutions.  Thus $H=E$, and also but
less importantly $D$, must vanish.
Similarly, from (2.31) one sees that $\pmb{P}$ must vanish, but this is not
surprising --- we expect static solutions to carry no momentum.  Note however:
angular momentum need not vanish for static configurations;
it can be constructed from
the current $m{\bf j} = \pmb{\cal P}$,
which in the static case must be divergence-free, according to (2.27).

Since $E$ can be given by (2.5) (provided there is sufficient regularity
so that the integral $\int d^2r\,\pmb{\nabla}\times {\bf j}$ vanishes)
we see that static solutions can exist only for $g\ge 1/m\kappa$.  Especially
interesting is the limiting case $g=1/m\kappa$, where the integrand is
non-negative and therefore must vanish on static solutions.  In this way the
$SO(2,1)$ conformal symmetry demands that {\it all\/} static solutions (at
$g=1/m\kappa$) satisfy
$$D\psi = 0 \eqno(2.36)$$
Together with the Chern--Simons constraint (1.26) this implies that $\rho$
satisfies the Liouville equation,
$$\nabla^2\ln\rho = - {2\over\kappa} \rho\eqno(2.37)$$
which can be integrated explicitly in terms of two arbitrary functions, which
are further specified by the physical requirements that one may wish to impose
on static solutions.$^{1,\,3}$

[In the non-Abelian case, the analogous equations, with $\psi$ in the same
fundamental representation as the gauge fields, realize a two-dimensional
reduction of four-dimensional self-dual gauge field equations in a space with
signature $(++--)$ and lead to many integrable systems, principally the Toda
system.$^4$]

As is well-known and was remarked in the Introduction,
the Liouville equation is invariant against conformal
redefinition of the two-dimensional plane. In Euclidean space this involves
the complex variable $x + iy =
z$ transforming into
an arbitrary function of $z$, but not of $z^*$:  Our
next task will be to understand the properties of the action (1.23) (at
$g=1/m\kappa)$ that are responsible for this infinite symmetry.  In fact its
stationary points are conformally invariant.

But before turning to this topic, we point out that the above described
transformations can be used to generate interesting new solutions from the
explicitly determined static solution.  First by Galileo [(2.10), (2.11)] or
conformal [(2.14), (2.15)] boosting of static solutions, one obtains time
dependent solutions to the Chern--Simons model.  Moreover, by performing
transformations with time-dependent parameters, one finds time-dependent
solutions to the Chern--Simons model with external, appropriately constructed
electric and magnetic fields as well as an external harmonic oscillator.$^7$
\goodbreak
\bigskip
\noindent{\bf B.\quad Infinite-Dimensional Symmetry Group of Stationary
Points of the Action}
\medskip
\nobreak
The dilation transformation (2.12) and (2.13) rescales the spatial coordinate
${\bf r}$.  Here we inquire about the response of the action (2.1a), (2.5) to
a conformal redefinition of spatial coordinates,
$${\bf r}' = {\bf r}'({\bf r}) \eqno(\hbox{2.38a})$$
where
$$x'+iy' \equiv z' = z'(z) \eqno(\hbox{2.38b})$$
and time is unchanged, $t'=t$.

Generalizing (2.13), we posit a field transformation law with a weight,
$$\psi'({\bf r}') = {\partial z^*\over \partial z'{}^*} \psi({\bf r})
\eqno(2.39)$$
which apart from a phase is the square root of the Jacobian, as in (2.13) and
(2.15), while the choice of phase is dictated by the Hamiltonian (2.5).
[Since time is not transformed, we suppress the time argument.]
This has the consequence that the density transforms with the
Jacobian as in (2.16).
$$\eqalignno{\rho'({\bf r}') &= J\rho({\bf r}) &(2.40) \cr
J &= \det\left\{{\partial r^i\over \partial r'{}^j}\right\}
= \left| {\partial z\over \partial z'}\right|^2 &(2.41)\cr}$$
For infinitesimal $\delta z = - f(z)$, this transformation law coincides with
(1.11b), taken in Euclidean space and $e^{\beta\varphi}$ identified with
$\rho$.  It further follows that the gauge potential transforms covariantly.
$$A'{}^i ({\bf r}') = A^j ({\bf r}) {\partial r^j\over \partial
r'{}^i}\eqno(2.42)$$
This is most easily proven by first
noting that ${\bf A}$, when given by Eq.~(1.25),
is transverse and satisfies $\pmb{\nabla}_{\bf r} \times {\bf A}({\bf r}) = -
{1\over\kappa} \rho({\bf r})$, and then verifying that ${\bf A}' ({\bf r}')$
in (2.42) also is transverse and satisfies
$\pmb{\nabla}_{{\rm r}'} \times {\bf A}' ({\bf
r}') = - {1\over\kappa} J \rho({\bf r}) = - {1\over\kappa}\rho'({\bf r}')$.
[In carrying out the differentiations it is useful to pass the complex
variables.]  It follows that
$D\psi \equiv \left( \partial_x - i \partial_y - i A^x + A^y\right)\psi$
transforms with the Jacobian.
$$D_{{\bf r}'} \psi'({\bf r}') = JD_{\bf r} \psi({\bf r}) \eqno(2.43)$$

So finally we can state the transformation law for the Lagrange density.
$$\eqalign{{\cal L} &= i \psi^* \partial_t \psi - {1\over 2m} \left|
D\psi\right|^2 + {1\over 2} \left( g - {1\over m\kappa}\right) \rho^2 \cr
&= i \psi^* \partial_t \psi - {\cal H}\cr}\eqno(2.44)$$
Evidently it is true that
$${\cal L}'({\bf r}') = J i \psi^*({\bf r}) \partial_t\psi({\bf r}) -
J^2 {\cal H} ({\bf r}) \eqno(2.45)$$
so that the Lagrangian transforms as
$$L' = \int d^2r'\,{\cal L}'({\bf r}') = \int d^2ri\psi^*({\bf r}) \partial_t
\psi({\bf r}) - \int d^2{\bf r}\, J {\cal H} ({\bf r}) \eqno(2.46)$$
One factor of the Jacobian disappears when changing spatial variables in the
integration, and the symplectic form $\int d^2r\, i \psi^* \partial_t \psi$
is invariant.  But the Hamiltonian density ${\cal H}$
remains with one factor $J$, hence the
total Lagrangian is not in general invariant, and neither is the action, the
time integral of $L$ --- because $t$ is not changed in the present
transformation rules
[in contrast to (2.12) and (2.14)].  [It does not appear possible to find a
transformation of time that would restore invariance.]

However, for static solutions we know that $E=\int d^2r\,{\cal H}$
vanishes.  If this vanishing is due to the local vanishing of ${\cal H}$, as
is true at $g = 1/m\kappa$, then the static critical points of the action are
invariant.  This then shows that static solutions with zero ${\cal H}$ will be
mapped into each other by spatial conformal transformations --- the dilation
(2.12) expands to an infinite symmetry group on the solutions, but there are
no new constants of motion.

[Since in the non-Abelian generalization, with matter in the adjoint
representation, equations are dimensional reductions of self-dual Yang--Mills
equations in four dimensions,$^4$ the {\it finite-dimensional\/}
conformal invariance of
the latter$^8$
is seen to survive the dimensional reduction, and in two dimensions
expands to the {\it infinite-dimensional\/} conformal group.]

On the other hand, in the effective field theories for the eikonal regime
(1.29), (1.31),
where there is no Hamiltonian to begin with, the transformations (2.38),
(2.39) {\it are\/} symmetries of
the action, and also $\tau$ may be arbitrarily reparametrized.  Note that owing
to the derivative relation (1.30) between $\Omega^\pm$ and $\psi$; $\psi =
\sqrt{2}{\partial\over\partial z^*}$
($\Omega^+-i\Omega^-$), the transformation law
for $\Omega^\pm$ is without the weight factor,
$$\Omega'{}^\pm ({\bf r}') = \Omega^\pm ({\bf r}) \eqno(2.47)$$
which arises for $\psi$, as in (2.39), when the derivative is taken.
\goodbreak
\bigskip
\noindent{\bf III.\quad CONCLUSION AND SUGGESTIONS FOR FURTHER RESEARCH}
\medskip
\nobreak
The rigid scale invariance of the action for non-relativistic
$(2+1)$-dimensional field theory with quartic self-interaction and coupling to
a Chern--Simons gauge field, expands at the static critical points of the
action to the infinite conformal group on the plane.  The scale symmetry
allows establishing the important result that static solutions carry zero
energy, and the infinite conformal symmetry ``explains'' why the static system
is completely integrable.  The kinetic action of effective eikonal field
theories also possesses the infinite symmetry.

The Chern--Simons
model at $g = 1/m\kappa$ is the bosonic partner of an $N=2$ supersymmetric
theory with fermions and the invariance of the extended action against the
supersymmetric generalization of the bosonic symmetries (2.6) -- (2.15) has
been
established.$^9$  While the invariances of the static critical points in the
supersymmetric action have not been explicitly checked, they too presumably
enjoy an infinite conformal symmetry, because the supersymmetric static
equations retain the form of the bosonic equations.

In our considerations, the possibility of quantum symmetry breaking anomalies
has been ignored.  It is known that the quartic self-interaction, which as we
have seen is formally scale invariant, suffers from quantum scale
anomalies.$^{10}$  This
is particularly clear in the first quantized framework, where the
two-dimensional $\delta$-function potential, while scaling classically as
$r^{-2}$, does not give rise to energy-independent phase shifts, as is
required by scale invariance and is explicitly realized by the scale invariant
$1/r^2$
potential.  There is a quantum scale anomaly --- the simplest example of the
anomaly phenomenon.$^{11}$  On the other hand, anomalies in the theory with
{\it
both\/} quartic self-coupling and Chern--Simons interaction have thus far not
been assessed; in fact there is some indication of anomaly
cancellation, even without
supersymmetry.$^{12}$  Further research on this question would be
interesting.$^{13}$
\vfill
\eject

\centerline{\bf REFERENCES}
\medskip
\item{1.}For a discussion of non-relativistic Chern--Simons theory
and its relation through second quantization to the particle mechanics of
(1.22), see {\it e.g.\/} R.~Jackiw and S.-Y.~Pi, {\it Phys. Rev. D\/} {\bf
42}, 3500 (1990).
\medskip
\item{2.}For gravity: H.~Verlinde and E.~Verlinde, {\it Nucl. Phys.\/} {\bf
B371}, 246 (1992);
for Maxwell theory: R. Jackiw, D. Kabat and M. Ortiz, {\it Phys. Lett. B\/} {
\bf 277}, 148 (1992).
\medskip
\item{3.}For Abelian Chern--Simons interactions: R.~Jackiw and S.-Y.~Pi,
{\it Phys. Rev. Lett.\/} {\bf 64}, 2969 (1990); (C) {\bf 66}, 2682 (1992) and
Ref.~[1].
\medskip
\item{4.}For non-Abelian Chern--Simons interactions: B.~Grossman, {\it
Phys. Rev. Lett.\/} {\bf 65}, 3230 (1990); G.~Dunne, R.~Jackiw, S.-Y.~Pi and
C.~Trugenberger, {\it Phys. Rev. D\/} {\bf 43}, 1332 (1991); G.~Dunne, {\it
Commun. Math. Phys.\/} (in press).
\medskip
\item{5.}S. Takagi, {\it Prog. Theor. Phys.\/} {\bf 84}, 1019 (1990), {\bf
85}, 463, 723 (1991), {\bf 86}, 783 (1991).
\medskip
\item{6.}D. Freedman and A. Newell (unpublished).
\medskip
\item{7.}Z. Ezawa, M. Hotta and Z. Iwazaki, {\it Phys. Rev. Lett.\/} {\bf 67},
441 (1991); {\it Phys. Rev. D\/} {\bf 44}, 452 (1991);
R.~Jackiw and S.-Y.~Pi, {\it Phys. Rev. Lett.\/} {\bf 67}, 415 (1991)
and {\it Phys. Rev. D\/} {\bf 44},  2524 (1991).
\medskip
\item{8.}R. Jackiw and C. Rebbi, {\it Phys. Rev. D\/} {\bf 14}, 517 (1977).
\medskip
\item{9.}M. Leblanc, G. Lozano and H. Min, {\it Ann. Phys.\/} (NY) (in press).
\medskip
\item{10.}O. Bergman, MIT preprint CTP\#2045 (1991).
\medskip
\item{11.}R. Jackiw, in {\it M. A. B. B\'eg Memorial Volume\/}, A. Ali and P.
Hoodbhoy, eds. (World Scientific, Singapore, 1991).
\medskip
\item{12.}G. Lozano, {\it Phys. Lett. B\/} (in press).
\medskip
\item{13.}O. Bergman, in preparation.
\par
\vfill
\end